\documentclass[fleqn,12pt,twoside]{article}
\usepackage[headings]{style}
\usepackage{amssymb}
\usepackage{amsmath}

\readRCS
$Id: espcrc1.tex,v 1.2 2004/02/24 11:22:11 spepping Exp $
\usepackage{graphicx}
\usepackage[figuresright]{rotating}

\title{Pressure of the standard model}

\author{A. Gynther\address[HIP]{Theoretical Physics Division, Department of Physical Sciences and Helsinki Institute of Physics, 
P.O. Box 64, FIN-00014 University of Helsinki, Finland}%
        \thanks{Present address: Department of Physics \& Astronomy, Brandon University, Brandon, MB R7A 6A9, Canada. E-mail: gynthera@brandonu.ca.}}
               
\runtitle{Pressure of the standard model}
\runauthor{A. Gynther}

\begin{document}

\maketitle

\begin{abstract}
We review the computation of the thermodynamic pressure of the entire minimal
standard model to three loop order, performed in \cite{AGpres}.
\end{abstract}

\section{INTRODUCTION}

Computing the thermodynamic pressure of the gauge field theories, defined by the path integral
\begin{equation}
p(T) = \lim_{V\rightarrow \infty} \frac{T}{V}\int\mathcal{D}\varphi\,\mathrm{exp}(-S_\mathrm{Euclidean}[\varphi]),
\end{equation}
has attracted a lot of interest over the years. In addition to being an interesting 
computation in itself, pressure is a phenomenologically important quantity when trying to understand the properties of 
quark-gluon plasma produced in heavy-ion collisions (relevant physics being described by QCD) and also in cosmology
(see, for example, \cite{HP}). In QCD with massless quarks, the expansion of the pressure in terms of the gauge 
coupling $g$ is known today to the last perturbatively computable term \cite{g6pres}. In the full standard model, the pressure
is less well known, the interest having been focused on the properties of the electroweak phase transition (crossover) which
depend only on the pressure difference between the phases of the theory. In this talk, we will review our recent computation
of the pressure of the full minimal standard model to three loop order, or to order $g^5$ in the coupling constants. We will
also outline some of the differences arising in the computations of the QCD pressure and the full standard model pressure.  

\section{FRAMEWORK OF THE COMPUTATION}

As well known, straightforward perturbative computations in gauge field theories at high temperatures are inhibited by infrared 
divergences that require resummation \cite{Linde}. A convenient way to perform the resummation is to construct effective field 
theories for the infrared sensitive modes (the bosonic zero modes) by integrating out the infrared safe modes 
(see \cite{AGpres,KKgen} for details). In case of the standard model, the resulting effective theory will be a three dimensional 
gauge field theory with fundamental (Higgs) and adjoint (temporal components of the gauge fields, corresponding to the electrostatic 
modes of the full theory) scalars, defined by the action\footnote{We neglect the QCD contribution here and consider just the electroweak
case.}
\begin{eqnarray}
S_\mathrm{E} & = & \int\mathrm{d}^3x \left(\frac{1}{4}F_{ij}F_{ij} + \frac{1}{4}G^a_{ij}G^a_{ij} + \frac{1}{2}(D_i A_0^a)^2
+ \frac{1}{2}(\partial_i B_0)^2 + \frac{1}{2}m_D^2A_0^aA_0^a + \frac{1}{2}m_{D}'^2B_0B_0 \right. \nonumber \\
& & \;\;\;\;\;\;\;\;\;\;\; \left. + |D_i\Phi|^2 + m_3^2|\Phi|^2 + \mathcal{L}_\mathrm{E,int}(\Phi,A_0^a,B_0)\right).
\end{eqnarray}
In this theory, the scalars have obtained thermal masses ($m_3^2 \sim -\nu^2 + g^2T^2$ for the Higgs and $m_D^2 \sim m_D'^2 \sim g^2T^2$ for
the adjoint scalars\footnote{Although the electroweak theory contains many different couplings, we use a power counting rule 
$g^2 \sim g'^2 \sim \lambda \sim g_Y^2$, $\nu^2 \lesssim g^2T^2$ to express all order of magnitude estimates in terms of
the single gauge coupling $g^2$.}) that serve to regulate (some of) the infrared divergences. 

Within this framework, assuming for now that the temperature is well above the temperatures corresponding to the electroweak 
crossover, the recipe for calculating the pressure is:
\begin{itemize}
\item Compute the strict perturbative expansion of the pressure in the full theory, schematically giving us (using dimensional
regularization, $d=4-2\epsilon$)
\begin{equation}
p_\mathrm{E}(T) \sim T^4\left[1+g^2+g^4\left(1/\epsilon + 1 + \ln \Lambda/T\right) + \mathcal{O}(g^6)\right],
\end{equation}
leaving uncancelled infrared divergences at order $g^4$ (three loop order).
\item Construct the three dimensional effective theory for the bosonic zero modes and compute its pressure, schematically
\begin{equation}
p_\mathrm{M}(T)/T \sim m^3 + g_3^2 m^2\left(1/\epsilon + 1 + \ln \Lambda/m\right) + g_3^4 m + \mathcal{O}(g_3^6).
\end{equation}
Here $m$ refers both to $m_3$ and $m_D$ and $g_3^2 = g^2T$. 
\end{itemize}
The total pressure is then given as the sum of these two contributions, 
$p(T) = p_\mathrm{E}(T) + p_\mathrm{M}(T) + \mathcal{O}(g^6)$. The $1/\epsilon$ poles in $p_\mathrm{M}(T)$ exactly cancel 
those of $p_\mathrm{E}(T)$, thus making $p(T)$ infrared finite.

\subsection{What happens near the electroweak crossover?}

The perturbative expansion of the QCD pressure becomes unreliable at temperatures near $T_c$ because confinement effects
become important ($\Lambda_\mathrm{QCD} \sim T_c$). The Landau pole of the electroweak theory, however, corresponds to a
length scale comparable to the radius of the Earth and one can therefore expect the confinement effects to be negligible
in the electroweak case, enabling us to extend the perturbative computations all the way down to the electroweak crossover.

Near the crossover region the fundamental scalar, which drives the transition, becomes light compared to the adjoint scalars,
thus introducing a new mass hierarchy to the system. Consequently, the computation outlined above
becomes invalid since it implicitly assumes that both the fundamental and the adjoint scalars can be considered heavy with masses 
$m\sim gT$. However, since the adjoint scalars are now heavy compared to the fundamental scalar, they can be
integrated out leading to a new effective theory describing just the fundamental scalar and the magnetic components of the 
gauge bosons,
\begin{equation}
S_\mathrm{E2} = \int\mathrm{d}^3x \left(\frac{1}{4}F_{ij}F_{ij} + \frac{1}{4}G^a_{ij}G^a_{ij} + |D_i\Phi|^2 + \widetilde{m}_3^2|\Phi|^2
+ \widetilde{\lambda}|\Phi|^4\right),
\end{equation}
where $\widetilde{m}_3^2 \sim m_3^2 - g_3^2m_D \ll m_D^2$ (valid near the crossover). The contribution of the effective theories 
to the pressure is then divided 
into two parts: contribution from the adjoint scalars, $p_\mathrm{M1}$, and the contribution from the remaining degrees of freedom, 
$p_\mathrm{M2}$. These can be 
written as $p_\mathrm{M}(T) = p_\mathrm{M1}(T) + p_\mathrm{M2}(T)$, where 
\begin{eqnarray}
p_\mathrm{M1}/T & \sim & m_D^3 + g_3^2m_D^2\left(1/\epsilon + 1 + \ln\Lambda/T\right) + g_3^4m_D\left(1/\epsilon + 1 + \ln\Lambda/T\right) + \mathcal{O}(g_3^6),\\
p_\mathrm{M2}/T & \sim & \widetilde{m}_3^3 + \widetilde{g}_3^2\widetilde{m}_3^2\left(1/\epsilon + 1 + \ln\Lambda/\widetilde{m}_3\right)
+ \mathcal{O}(\widetilde{g}_3^4\widetilde{m}_3).
\end{eqnarray}
Since the fundamental scalar becomes light near the crossover, new infrared divergences (terms $\sim g_3^4m_D/\epsilon$) appear that 
cancel only when the contribution from the second effective theory $S_\mathrm{E2}$ is taken into account. It is noteworthy that the new 
poles appear at order $g^5$, thus leading to terms of the order of $g^5\ln g$ in the expansion of the full pressure. Such terms are not 
present in the expansion of the QCD pressure.

\section{NUMERICAL RESULTS}

\begin{figure}[!t]
\begin{minipage}{75mm}
\includegraphics*[width=75mm]{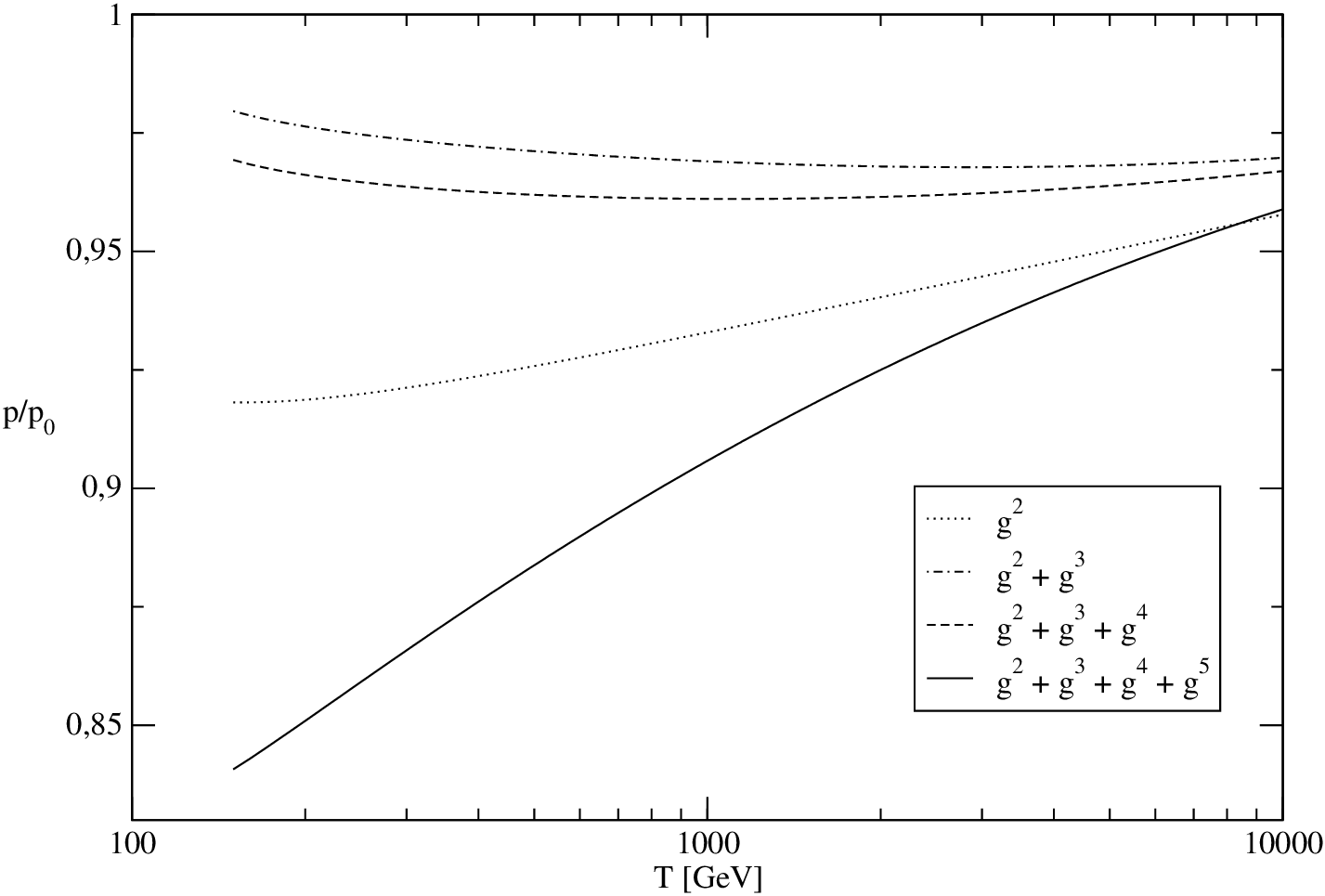}
\caption{The expansion of the pressure of the minimal standard model to various orders in $g^2$.} \label{fig:pressure}
\end{minipage}
\hfill
\begin{minipage}{75mm}
\includegraphics*[width=75mm]{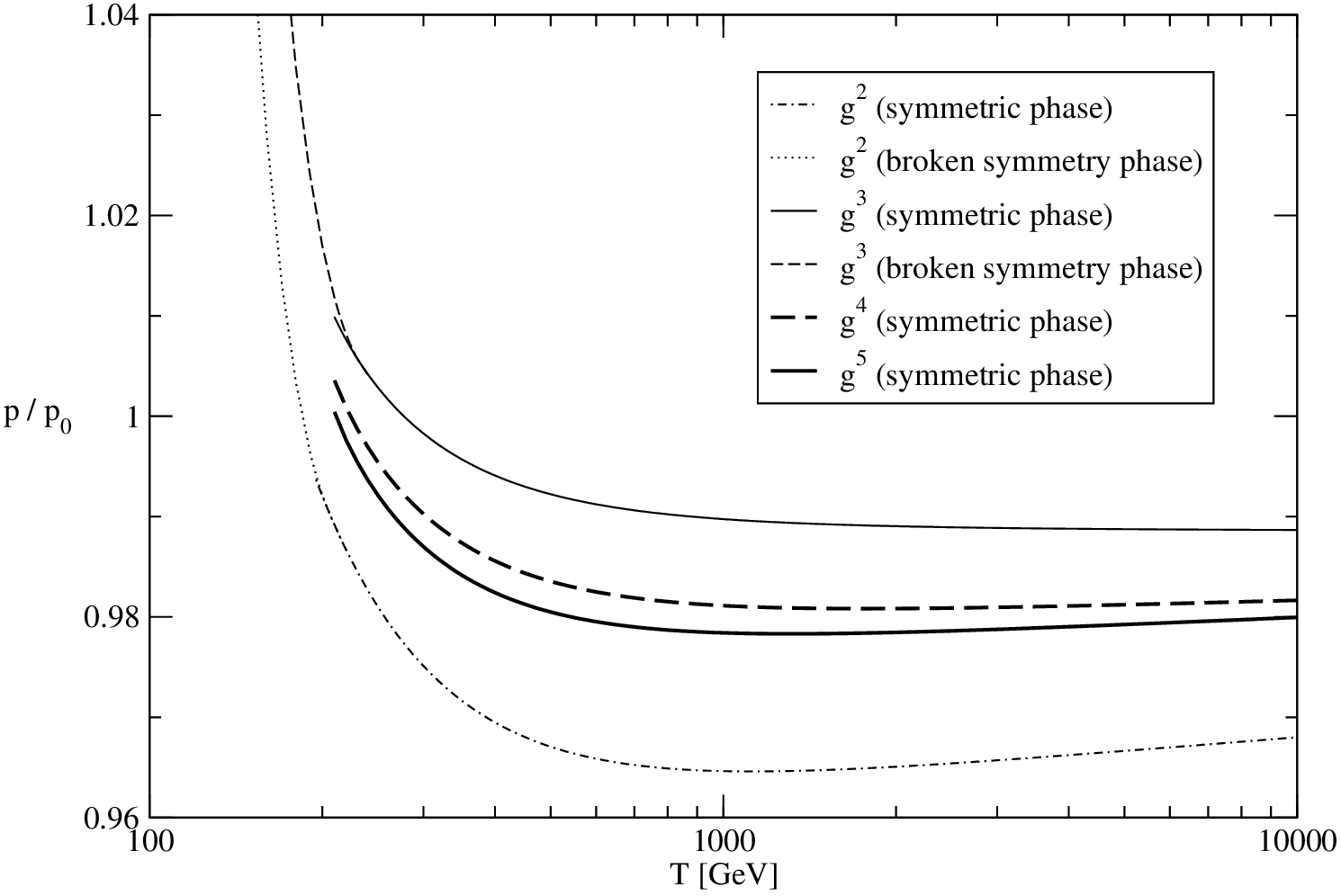}
\caption{The expansion of the pressure of the SU(2) + Higgs theory to various orders in $g^2$.} \label{fig:pressure2}
\end{minipage}
\end{figure}

The detailed analytic expression of the expansion of the pressure is given in \cite{AGpres} and we will here content to reviewing some
numerical results obtained from that. In Fig.~\ref{fig:pressure} we have plotted the expansion at different orders in $g^2$, normalized 
to the pressure of ideal gas of massless particles $p_0(T) = 106.75\cdot \pi^2T^4/90$. The result deviates strongly from the ideal gas
result, up to as much as 15\%. As discussed in \cite{HP}, such a departure from the ideal gas result may affect the predictions concerning
the WIMP relic density in the universe.

As can also be seen from the plot, the expansion does not converge well; the order $g^5$ term produces a significant correction to the 
pressure compared to the previous terms. The poor convergence is mostly due to the QCD contributions (of the 106.75 bosonic degrees of 
freedom in the standard model, 79 come from QCD). To analyze the convergence within the electroweak sector, we can study a simpler 
SU(2) + Higgs theory. As can be seen from the plot in Fig.~\ref{fig:pressure2}, also normalized to the ideal gas pressure, the expansion 
within just the electroweak sector converges better: each new term in the expansion is smaller than the previous one. Similar conclusions
can be reached by studying the scale dependence of the expansion.

\begin{figure}[!t]
\begin{centering}
\includegraphics[width=75mm]{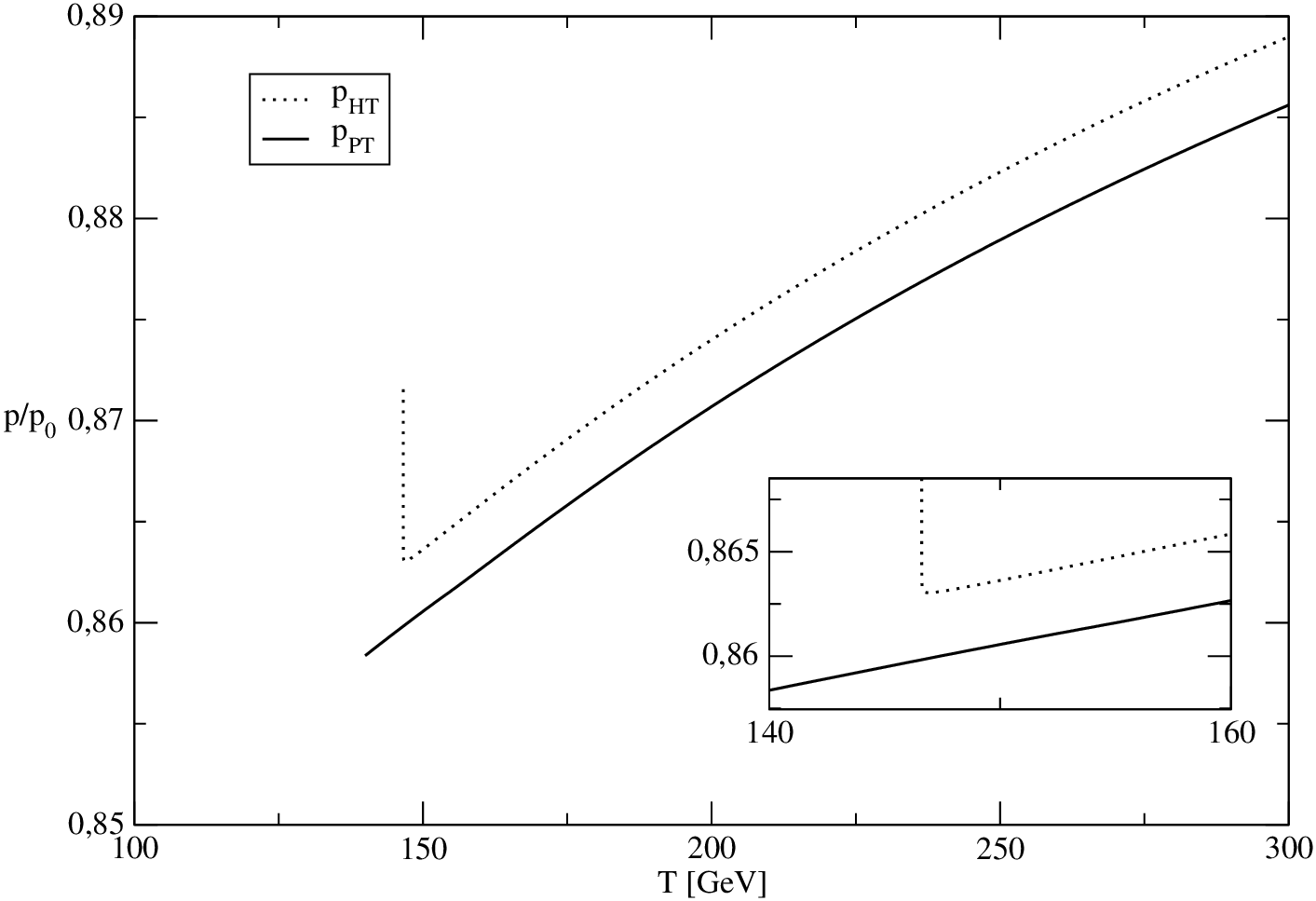}
\caption{The pressure near the electroweak crossover according to the high temperature computation ($p_\mathrm{HT}(T)$) and the 
computation taking into account that the fundamental scalar is light near the crossover.} \label{fig:crossover}
\end{centering}
\end{figure}

As we approach the crossover region, the high temperature computation ceases to be valid as discussed in the previous section. This
can be explicitly seen in Fig.~\ref{fig:crossover}. The high temperature expansion of the pressure (denoted by $p_\mathrm{HT}(T)$ 
in the plot) contains an unphysical singularity due to the infrared divergences which appear when $m_3\rightarrow 0$. Taking into account
that the Higgs scalar becomes light near the crossover, the expansion behaves well all the way down to the electroweak crossover 
($p_\mathrm{PT}(T)$). This result can therefore be used for a precise determination of the expansion rate of the universe near the
electroweak crossover, which is important in many cosmological studies. 

This work has been done in collaboration with M.~Veps\"al\"ainen.

\end{document}